\documentclass[a4paper,aps,english,twocolumn,floatfix,amsfonts,amssymb,superscriptaddress]{revtex4}
\usepackage{epstopdf}
\usepackage{graphicx}
\usepackage{dcolumn}
\usepackage{bm}
\usepackage{bbm,bbold}
\usepackage{color}
\usepackage{xcolor, soul}
\sethlcolor{green}
\usepackage{amsfonts}
\usepackage{amsmath}
\usepackage{mdframed}
\usepackage[normalem]{ulem}
\usepackage{mathrsfs}   
\usepackage[percent]{overpic}
\usepackage[colorlinks=true,citecolor=blue]{hyperref}
\hypersetup{colorlinks=true,citecolor=blue,linkcolor=blue,urlcolor=blue}
\DeclareRobustCommand{\rchi}{{\mathpalette\irchi\relax}}
\newcommand{\irchi}[2]{\raisebox{\depth}{$#1\chi$}} % inner command, used by \rchi
\begin{document}
\title{Nonlinear anomalous photocurrents in Weyl semimetals}

\author{Habib Rostami}
\email{habib.rstm@gmail.com}
\affiliation{Istituto Italiano di Tecnologia, Graphene Labs, Via Morego 30, I-16163 Genova,~Italy}

\author{Marco Polini}
\affiliation{Istituto Italiano di Tecnologia, Graphene Labs, Via Morego 30, I-16163 Genova,~Italy}
\begin{abstract}
We study the second-order nonlinear optical response of a Weyl semimetal (WSM), i.e. a three-dimensional metal with linear band touchings acting as point-like sources of Berry curvature in momentum space, termed ``Weyl-Berry monopoles". We first show that the anomalous second-order photocurrent of WSMs can be elegantly parametrized in terms of Weyl-Berry dipole and quadrupole moments. We then calculate the corresponding charge and node conductivities of WSMs with either broken time-reversal invariance or inversion symmetry. In particular, we predict a 
%universal 
dissipationless second-order anomalous node conductivity for WSMs belonging to the TaAs family. 
\end{abstract}
\maketitle
\section{Introduction} 
Recent years have witnessed momentous interest in the linear response of topological materials to external probes~\cite{Xiao_rmp_2010,Nagaosa_rmp_2010}. In particular, many crystals harbor hot spots of non-vanishing Berry curvature in the Brillouin zone (BZ)~\cite{Xiao_rmp_2010,Nagaosa_rmp_2010}, whose existence is guaranteed by either broken inversion symmetry or broken time-reversal symmetry. In the latter case, the crystal displays the anomalous Hall effect~\cite{Nagaosa_rmp_2010}. In the former case, it has been shown, 
that two-dimensional crystals with broken inversion symmetry manifest the valley Hall effect~\cite{Xiao_prl_2007,dixiao_prl_2012,gorbachev_science_2014,lensky_prl_2015,sui_naturephys_2015,shimazaki_naturephys_2015,mak_science_2014,lee_naturenano_2016,beconcini_prb_2016}. In this case, no net charge flows in response to an electric field, but charge-neutral 
transverse valley currents propagate microns aways from the current injection path yielding large non-local electrical signals~\cite{gorbachev_science_2014,lensky_prl_2015,sui_naturephys_2015,shimazaki_naturephys_2015,beconcini_prb_2016}.

These concepts have been introduced in the context of linear response theory~\cite{Giuliani_and_Vignale} and, to the best of our knowledge, existing experiments probing topological effects have mainly been carried out in the linear response regime. It is therefore natural to try and understand what are the implications of a topological band structure on {\it nonlinear} transport and optics. A few pioneering works in this direction have recently appeared in the literature~\cite{Spivak_arxiv_2009,Sodemann_prl_2015,Morimoto_SA_2016,Morimoto_prb_2016,Cortijo_prb_2016,juan_arxiv_2016,chan_prb_2017,Wu_NaturePhys_2017,Konig_arxiv_2017,Golub_arxiv_2017,Ma_arxiv_2017}.

One of the simplest possible nonlinear optical effects occurring in a solid is that of rectified currents, also known as dc photocurrents or photogalvanic currents~\cite{Belinicher_1978,Ivchenko_Pikus_1978,Aversa_prb_1995,Sipe_prb_2000}. In nonlinear optics~\cite{Boyd}, indeed, a dc current can appear in response to an oscillating electric field, when one analyzes the response of a crystal up to (at least) second order in the applied field. At this order of perturbation theory, dc photocurrents are the result of ``injection''~\cite{Belinicher_1978,Ivchenko_Pikus_1978,Aversa_prb_1995,Sipe_prb_2000}, ``shift''~\cite{Belinicher_1978,Ivchenko_Pikus_1978,Sipe_prb_2000}, and ``anomalous''~\cite{Spivak_arxiv_2009,Sodemann_prl_2015} contributions. Interestingly, all of these three contributions may reveal subtle topological effects. In the literature, the injection current contribution is often termed ``circular photocurrent'' because a) it can only occur in response to circularly polarized light and b) it flips sign when the helicity of light changes sign.

The anomalous contribution to the dc photocurrent is called so because it stems from the anomalous velocity~\cite{anomalous_velocity,Sundaram_Niu}, which is controlled by Berry curvature. As we demonstrate below, in the clean limit also the anomalous contribution yields a circular photocurrent with properties a) and b).
\begin{figure}[t]
\centering
\includegraphics[width=85mm]{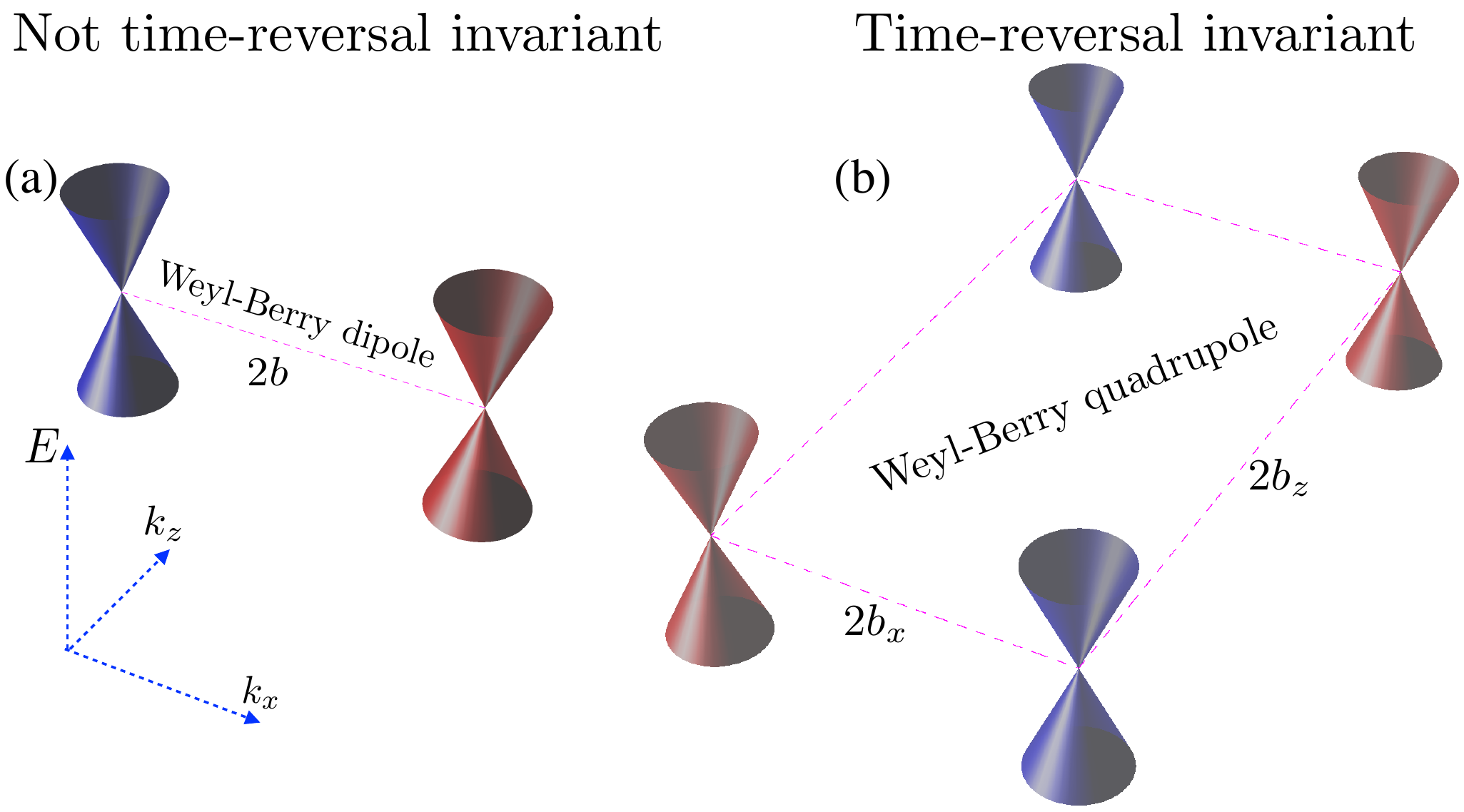}
\caption{(Color online) The simplest possible distributions of momentum-space Weyl-Berry monopoles in a 3D Weyl semimetal. The cones represent the low-energy band structure plotted as a function of $k_{x}$ and $k_{z}$, for $k_{y}=0$. Each crossing  (Weyl node) acts as a ``Weyl-Berry monopole'' in momentum space, i.e.~a point-like source of Berry curvature. The color code indicates the chiral charge $\rchi = \pm 1$ of each node. Panels (a) and (b) refer to crystals with a finite Weyl-Berry dipole and quadrupole, respectively. In the former case, time-reversal symmetry is broken. In the latter, inversion symmetry is broken. Note that WSMs with broken inversion symmetry come with a minimum of four nodes. \label{fig:wsm}}
\end{figure}

In this work we analyze the anomalous current contribution to the dc photocurrent in Weyl semimetals (WSMs). 
These are recently discovered materials~\cite{Hosur_2013,Yan_2017,Hasan_2017,Burkov_2017,Vishwanath_arxiv_2017} with robust linear band crossings in the BZ. Each crossing is usually called ``Weyl node'' and can be identified by its chirality~\cite{Hosur_2013,Yan_2017,Hasan_2017,Burkov_2017,Vishwanath_arxiv_2017}. Weyl nodes act as monopoles in momentum space~\cite{Hosur_2013,Yan_2017,Hasan_2017,Burkov_2017,Vishwanath_arxiv_2017}. Indeed, a Weyl node located at ${\bm k} = {\bm 0}$ in the BZ produces a Berry curvature that obeys the following relation: $(\partial/\partial {\bm k}) \cdot {\bm \Omega}^{(\rchi)}_{\pm}({\bm k}) = \pm 2\pi \rchi \delta^{(3)}({\bm k})$, where $\pm$ denotes conduction/valence band states and $\rchi=\pm$ is the Weyl-node chirality. Similarly to the case of a point-like electrical charge, this implies a power-law-decaying Berry curvature ${\bm \Omega}^{(\rchi)}_{\pm}({\bm k})= \pm \rchi {\bm k}/(2 k^{3})$, with $\rchi$ playing the role of an effective charge. In crystals with either broken time-reversal symmetry or broken inversion symmetry, Weyl nodes comes always in pairs with opposite chirality~\cite{Hosur_2013,Yan_2017,Hasan_2017,Burkov_2017,Vishwanath_arxiv_2017}. The total effective charge of the associated distribution of Weyl-Berry monopoles in the BZ is therefore zero. However, WSMs can carry a discrete distribution of Weyl-Berry monopoles with an associated non-vanishing higher-order multipole, such as a {\it dipole} or a {\it quadrupole} moment---see Fig.~\ref{fig:wsm}. Here, we are interested in type-I WSMs~\cite{Hosur_2013,Yan_2017,Hasan_2017,Burkov_2017,Vishwanath_arxiv_2017} where only electrons or hole pockets exist at the Fermi energy. In centrosymmetric WSMs with broken time-reversal symmetry %(BTRS)
~\cite{Wan_prb_2011,Hirschberger_nm_2016,Wang_prl_2016}, for example, a non-vanishing Weyl-Berry dipole exists in the BZ. The simplest case is that in which $N=2$ Weyl nodes~\cite{Yan_2017,McCormick_prb_2017} separated by a vector $2{\bm b}$ occur in the BZ---see Fig.~\ref{fig:wsm}(a). In such a system, an anomalous Hall current $\propto {\bm b} \times \bm{\mathcal E}$~\cite{Burkov_prl_2011,Goswami_prb_2013,Pesin_arxiv_2017} flows transversally to the applied electric field $\bm{\mathcal E}$ in the linear-response regime. On the other hand, in (time-reversal-invariant) WSMs with broken inversion symmetry %(BIS)
~\cite{Xu_science_2015,lv_prx_2015,Lv_np_2015,Yang_np_2015,Xu_np_2015,huang_nc_2015}, the first non-vanishing multipole is the Weyl-Berry  quadrupole. The simplest example is that of a WSM with $N=4$ Weyl nodes~\cite{Yan_2017,McCormick_prb_2017} in the BZ---see Fig.~\ref{fig:wsm}(b).

Theoretically, the topological properties of the injection-current-rate contribution to the circular photocurrent in WSMs have been highlighted in Refs.~\cite{juan_arxiv_2016,chan_prb_2017,Konig_arxiv_2017,Golub_arxiv_2017}. Experimentally, a very strong second-harmonic generation signal has been reported in WSMs with broken inversion symmetry~\cite{Wu_NaturePhys_2017}. Very recently, the photocurrent response of TaAs to circularly polarized mid-infrared light has been measured in Ref.~\cite{Ma_arxiv_2017}. We now proceed to highlight the topological and geometrical features of the anomalous contribution to dc photocurrents in type-I WSMs.

\section{Theory and Method} %Nonlinear response theory
In this work we are interested in dc photocurrents in response to light with frequency $\omega\tau\gg 1$, where $\tau$ is the shortest time scale associated with extrinsic effects (scattering of electrons against disorder, phonons, etc). The Hamiltonian we set out to study is $\hat {\cal H}(t)= \hat{\cal H}_0 - \hat {\bm P} \cdot \bm{\mathcal E}(t)$. The first term ($\hat{\cal H}_0$) is the WSM  Hamiltonian, while the second term describes the interaction of electrons with light, as described by a uniform time-dependent electric field $\bm{\mathcal E}(t)$. The electric polarization (electronic dipole moment per unit volume) operator is 
(see Appendix~\ref{app:pol}) 
\begin{equation}
 \hat {\bm P}=e \sum_{n, {\bm k}} \hat a^\dagger_{n}({\bm k})
 \left [i\frac{\partial  \hat a_{n}({\bm k}) }{\partial {\bm k}}+ \sum_{m} {\cal \bm A}_{nm}({\bm k}) \hat a_{m}({\bm k})\right ]~.
\end{equation}
Here, $\hat{a}^\dagger_{n}({\bm k})$ ($\hat{a}_{n}({\bm k})$) creates (annihilates) an electron in the Bloch state $|u_{n{\bm k}}\rangle$ with energy $E_{n}({\bm k})$ and 
$\bm{\mathcal A}_{nm}({\bm k})  \equiv  i   \langle u_{n{\bm k}}   |  {\partial u_{m{\bm k}}}/{\partial {\bm k}}\rangle$.  Notice that $\sum_{\bm k} \equiv  \int d^3 {\bm k}/(2\pi)^3$. 

The macroscopic polarization ${\bm P}(t)=  \langle \Psi_{0}|\hat{\bm P}|\Psi_{0}\rangle \equiv \langle\hat{\bm P}\rangle$ is obtained from the average of $\hat {\bm P}$ over the ground state $|\Psi_0 \rangle$. The time dependence originates from the quantum equation of motion. From the macroscopic polarization, one can calculate the macroscopic charge current, ${\bm J}(t)= \partial{\bm P}(t)/\partial t$, i.e.
\begin{equation}\label{eq:J_dpdt}
{\bm J}(t)= e \sum_{n, m, {\bm k}}  \bm{\mathcal  A}_{nm}({\bm k}) \frac{\partial \rho_{nm}({\bm k})}{\partial t} +e \sum_{n, {\bm k}}  \frac{\partial {\bm \zeta}_{nn}({\bm k})}{\partial t}~,
\end{equation}
where  $\rho_{nm}({\bm k}) \equiv   \langle \hat a^\dagger_{n}({\bm k}) \hat a_{m} ({\bm k})  \rangle$ are the elements of the density matrix and
${\bm \zeta}_{nm}({\bm k}) \equiv i   \langle \hat a^\dagger_{n}({\bm k})  {\partial \hat a_{m} ({\bm k}) }/{\partial {\bm k}}\rangle$. Equation~(\ref{eq:J_dpdt}) is non-perturbative in the strength of the applied electric field $\bm{\mathcal E}(t)$. 
We note that, at the equilibrium and in the absence of an external electric field,  $\rho_{nm}({\bm k})=\delta_{nm} n_{\rm F}(E_{n}({\bm k})) $ and
$2{\bm \zeta}_{nm}({\bm k}) = \delta_{nm} {\partial  n_{\rm F}(E_{n}({\bm k}))  }/{\partial {\bm k}} $, where $n_{\rm F}(E_{n}({\bm k}))$ is the Fermi distribution function. 
After lengthy but straightforward algebra (see Appendix~\ref{app:current}), 
we find that the charge current can be written as the sum of intra- and inter-band contributions, i.e.~${\bm J}(t)={\bm J}^{\rm intra}(t)+{\bm J}^{\rm inter}(t)$, with
 \begin{eqnarray}\label{eq:j_intra}
{\bm J}^{\rm intra}(t) &=& 
- \frac{e^2}{\hbar}  \bm{\mathcal E}(t) \cdot \sum_{n, {\bm k}}   \frac{\partial \bm{\mathcal R}_{n}({\bm k})  }{\partial {\bm k}} 
+\sum_{n, {\bm k}} \frac{e}{\hbar}\frac{\partial E_{n}({\bm k})  }{\partial {\bm k}} \rho_{nn}
\nonumber \\
&-&\frac{e^2}{\hbar}  \bm{\mathcal E}(t) \times  \sum_{n, {\bm k}}  {\bm \Omega}_{n}({\bm k}) \rho_{nn}
\end{eqnarray}
and
 \begin{eqnarray}\label{eq:j_inter}
{\bm J}^{\rm inter}(t) &=& -  \frac{e^2}{\hbar} \sum_{n\neq m, {\bm k}} \rho_{n m}\hat{\bm{\mathcal D}}_{mn}({\bm k})  \bm{\mathcal E}(t) \cdot   \bm{\mathcal A}_{nm}({\bm k})   
\nonumber \\
&+& e \sum_{n\neq m, {\bm k}} \bm{\mathcal A}_{nm}({\bm k})   \frac{\partial \rho_{nm}}{\partial t}~.
\end{eqnarray}
Here, ${\bm \Omega}_{n}({\bm k}) =  {\partial }/{\partial {\bm k}} \times \bm{\mathcal A}_{nn}({\bm k})$ is the Berry curvature~\cite{Xiao_rmp_2010} and 
$\bm{\mathcal R}_{n}({\bm k})=\bm{\mathcal A}_{nn}({\bm k})\rho_{nn}({\bm k})+  {\bm \zeta}_{nn}({\bm k})$.  In Eq.~(\ref{eq:j_inter}) we have introduced the generalized derivative
$\hat{\bm{\mathcal D}}_{nm}({\bm k}) = {\partial }/{\partial {\bm k}}  
+ i [\bm{\mathcal A}_{nn}({\bm k}) - \bm{\mathcal A}_{m m}({\bm k})]$. 
With the aid of the non-perturbative results (\ref{eq:j_intra})-(\ref{eq:j_inter}) one can study all nonlinear effects in arbitrary crystals, in the non-interacting limit~\cite{footnote}. In second-order perturbation theory, the second term in Eq.~(\ref{eq:j_intra}) contributes to the injection-current-rate~\cite{Sipe_prb_2000}. At the same order, the first term in the right-hand side of Eq.~(\ref{eq:j_inter}) contributes to the shift current, which will be the subject of another work.  Finally, the contribution to the photocurrent due to the second term in the right-hand side of Eq.~(\ref{eq:j_inter}) vanishes.

In this work, we focus on the anomalous photocurrent, i.e.~the third term in Eq.~(\ref{eq:j_intra}). 
We find that the anomalous contribution to the second-order current reads as following: (see Appendix~\ref{app:eq5}) 
\begin{align}\label{eq:j2_1}
&{\bm J}^{(2)}(\omega_\Sigma) = - i\frac{e^3}{2\hbar^2}  \int \frac{d^{3} {\bm k}}{(2\pi)^3} 
\sum_{n}n_{\rm F}(E_{n}({\bm k})) \nonumber \\ 
&\times
  \left \{   \frac{ \bm{\mathcal E}(\omega_{1})}{\omega_{1}}  
\cdot  \frac{\partial   }{\partial  {\bm k}}  
 \left [ \bm{\mathcal E}(\omega_2) \times  {\bm \Omega}_{n}({\bm k})\right ] 
 + \omega_1 \leftrightarrow \omega_2
  \right \},  
\end{align}
where $\omega_{\Sigma}\equiv \omega_1+\omega_2$, $\bm{\mathcal E}(\omega)$ is the Fourier transform of a single-frequency external electric field. Eq.~(\ref{eq:j2_1}), which is the first important result of this work, describes the anomalous contribution to all second-order nonlinear optical effects, like dc photocurrents, second-harmonic generation, sum- and difference-frequency phenomena, etc. In writing Eq.~(\ref{eq:j2_1}) we have enforced the {\it intrinsic permutation symmetry}~\cite{Butcher_and_Cotter}---i.e. the symmetry under the permutation $\omega_1 \leftrightarrow \omega_2$.

The second-order anomalous contribution to the dc  photocurrent can be calculated by setting $\omega_1=-\omega_2=\omega>0$ and $\omega_{\Sigma}=0$ in Eq.~(\ref{eq:j2_1}). Straightforward algebraic manipulations yield (see Appendix~\ref{app:eq6})
\begin{align}\label{eq:j2_3}
{\bm J}^{(2)}(0) &= \frac{e^3}{2h^2 \omega}\int \frac{d^3 {\bm k}}{2\pi} 
  \sum_{n} n_{\rm F}(E_{n}({\bm k})) 
\bigg [
\bm{\mathcal F}(\omega)
\frac{\partial}{\partial {\bm k}} \cdot { \bm \Omega}_{n}({\bm k}) 
\nonumber\\
&
- \bm{\mathcal F}(\omega) \cdot \frac{\partial { \bm \Omega}_{n}({\bm k})}{\partial {\bm k}}   
-\bm{\mathcal F}(\omega) \times  \left( \frac{\partial  }{\partial {\bm k}} \times { \bm \Omega}_{n}({\bm k})   \right)
\bigg ],
\end{align}
where $\bm{\mathcal F}(\omega)= i \bm{\mathcal E}(\omega)\times \bm{\mathcal E}^\ast(\omega)$ is a real-valued field oriented along the direction of propagation of light. 
Eq.~(\ref{eq:j2_3}) is the second important result of this work. It yields a vanishing anomalous second-order current for a linearly-polarized external field. (In this case indeed,  
$\bm{\mathcal E}(\omega) \parallel \bm{\mathcal E}^\ast(\omega)$ which implies $\bm{\mathcal F}(\omega)={\bm 0}$.) We conclude that a second-order anomalous current can emerge only in response to a circularly-polarized laser, in analogy to the injection-current-rate contribution~\cite{Sipe_prb_2000,Dyakonov,juan_arxiv_2016}.  
 The above relation is in contrast with the intrinsic quantum nonlinear Hall current introduced in Ref.~\cite{Sodemann_prl_2015} because it predicts a vanishing second-order anomalous current in response to a linearly
polarized electric field. We believe that this inconsistency originates from the intrinsic permutation symmetry~\cite{Butcher_and_Cotter}, which we have correctly taken it into account while it was missed by the authors of Ref.~\cite{Sodemann_prl_2015}. 

Note that Eq.~(\ref{eq:j2_3}) is well-defined even in the absence of extrinsic effects (e.g.~disorder). In this sense, it conceptually differs from conventional circular photocurrents derived from injection-current-rate-type contributions~\cite{juan_arxiv_2016,chan_prb_2017,Belinicher_1978,Ivchenko_Pikus_1978,Sipe_prb_2000,Ma_arxiv_2017,Konig_arxiv_2017,Golub_arxiv_2017}, which are proportional to the momentum relaxation time, $\tau$,~\cite{Dyakonov}. 
Therefore, one can roughly write $ |{\bm J}^{(2)}_{\rm injection}(0) |/ |{\bm J}^{(2)}_{\rm anomalous}(0)| \sim \omega\tau $
where $\tau$ stands for the relaxation time and $\omega$ is the driving field (incident laser) frequency. This relation implies that by considering a finite value of $\tau$, the anomalous dc-current dominates in the low-frequency regime, while at high frequency the injection current is the leading contribution. In any case, the inclusion of disorder is well beyond the scope of our Article. 
Finally, Eq.~(\ref{eq:j2_3}) reveals very interesting topological and geometrical features. It depends only on the momentum-space divergence, curl, and gradient of the Berry curvature, ${\bm \Omega}_{n}({\bm k})$. 
In time-reversal symmetric systems, the  anomalous charge current vanishes at the linear response level because of the {\it odd} symmetry of the Berry curvature under time reversal. This implies that the second-order anomalous dc current is the leading contribution because all derivatives of Berry curvature with respect to momentum are {\it even} functions of ${\bm k}$.
\section{Theory of  anomalous circular photocurrents in WSMs} 
We start by evaluating the single-monopole contribution to the anomalous dc photocurrent at zero temperature. For a single Weyl-Berry monopole located at ${\bm k}={\bm b}$, the Berry curvature field is give by ${\bm \Omega}^{(\rchi)}_{n=\pm}({\bm k})=n\rchi ({\bm k}-{\bm b})/2|{\bm k}-{\bm b}|^3$.  
The curl of the Berry curvature vanishes, i.e.~${\partial  }/{\partial {\bm k}} \times {\bm \Omega}^{(\rchi)}_{n}({\bm k})=0$. On the other hand, as repeatedly emphasized earlier, its divergence is a topological quantity that depends on the effective charge $\rchi$: 
\begin{align}\label{eq:integral_div}
\int \frac{d^{3} {\bm k}}{2\pi} \frac{\partial   }{\partial  {\bm k} }  \cdot {\bm \Omega}^{(\rchi)}_{n}({\bm k})  n_{\rm F}(E_{n}({\bm k}))= n \rchi f_n(k^{(\rchi)}_{\rm F})~,
\end{align}
where $n=\pm1$, $E_{n}({\bm k}) = \pm \hbar v |{\bm k}-{\bm b}|$, $k^{(\rchi)}_{\rm F}$ is the node Fermi wave number, and $f_{n}(x)\equiv (1-n)/2+n\Theta(x)$ with $\Theta(x > 0)=1$ and $\Theta(x \le 0)=0$. In the definition of conduction/valence band energies $E_{\pm}({\bm k})$, $v$ is the Weyl fermion's velocity.

Finally, also the gradient of ${\bm \Omega}^{(\rchi)}_{n}({\bm k})$ contains a topological term related to the effective charge:
\begin{align}\label{eq:integral_grad}
\int \frac{d^{3} {\bm k}}{2\pi} \frac{\partial  {\bm \Omega}^{(\rchi)}_{n}({\bm k})}{\partial  {\bm k} }   n_{\rm F}(E_{n}({\bm k})) 
&=  n \rchi \bigg [ \frac{f_n(k^{(\rchi)}_{\rm F})}{3} \mathbb{1} 
\nonumber\\
&
- \bm{\mathcal M}_{n}({\bm b},k^{(\rchi)}_{\rm F}) \bigg]~, 
\end{align}
where $\mathbb{1}$ is the $3\times3$ identity matrix and $\bm{\mathcal M}_{n}({\bm b},k^{(\rchi)}_{\rm F})$ is a traceless rank-two tensor that depends on $k^{(\rchi)}_{\rm F}$ and the direction of ${\bm b}$. Its components are defined by
\begin{align}
& {\mathcal M}_{n,\alpha\beta}({\bm b},k^{(\rchi)}_{\rm F}) =  \int  \frac{d^{3} {\bm k} }{4\pi} n_{\rm F}(E_{n}({\bm k}))
\nonumber \\ 
&\hspace{10mm}\times \frac{3 ({\bm k}-{\bm b})_\alpha ({\bm k}-{\bm b})_\beta -|{\bm k}-{\bm b}|^2 \delta_{\alpha\beta}}{|{\bm k}-{\bm b}|^5}~.
\end{align}
We now introduce the single-Weyl-node second-order anomalous conductivity $\sigma^{(2)}_{(\rchi,{\bm b}); \alpha\beta}(\omega)$ in such a way that the single-node dc photocurrent is given by
$J^{(2)}_{(\rchi,{\bm b}); \alpha}(0)  = \sum_{\beta} \sigma^{(2)}_{(\rchi,{\bm b}); \alpha\beta}(\omega) {\cal F}_{\beta}(\omega)$.
We find
\begin{equation}\label{eq:single-node-second-order}
{\bm \sigma}^{(2)}_{(\rchi,{\bm b})}(\omega) =   \frac{ \rchi e^3}{2 h^2\omega } \bigg \{ \frac{2[2\Theta(k^{(\rchi)}_{\rm F})-1]}{3} \mathbb{1}+   {\bm{\mathcal M}}_{\rm cv}({\bm b},k^{(\rchi)}_{\rm F}) \bigg\}
\end{equation}
with $\bm{\mathcal M}_{\rm cv}({\bm b},k^{(\rchi)}_{\rm F}) \equiv  {\bm {\mathcal M}}_{+}({\bm b},k^{(\rchi)}_{\rm F})- {\bm {\mathcal M}}_{-}({\bm b},0)$ for an electron-doped WSM. For a hole-doped WSM one has an identical final result for $\bm{\mathcal M}_{\rm cv}$ due to particle-hole symmetry. We note that the single-node second-order conductivity (\ref{eq:single-node-second-order}) is proportional to $\chi$ and therefore topological in nature. This feature is protected as long as inter-node scattering is neglegible~\cite{Franz_Molenkamp}. The tensor $\bm{\mathcal M}_{\rm cv}$ brings in a geometrical dependence. It is possible to show that for a Weyl-Berry monopole at the origin, i.e.~for ${\bm b}= {\bm 0}$, ${\bm {\mathcal M}}_{\rm cv}({\bm 0},k^{(\rchi)}_{\rm F})={\bm 0}$. This implies that the definition of the $\bm{\mathcal M}_{\rm cv}({\bm b},k^{(\rchi)}_{\rm F})$ tensor is ambiguous as it depends on the location of the origin. This ambiguity, which is well-known in electrostatics when one defines electrical multipoles, is lifted when one considers a discrete distribution of chiral monopoles, as we show later.

For a single Weyl node located at an arbitrary position (except for the origin) in the $\hat{\bm k}_{x}$-$\hat{\bm k}_{z}$ plane, i.e.~at $ {\bm b} = b[\hat{\bm k}_{x}\cos(\phi) + \hat{\bm k}_{z} \sin(\phi)]$, we find (see Appendix~\ref{app:eq11})  
\begin{align}\label{eq:M}
\bm{\mathcal M}_{\rm cv}&({\bm b},k^{(\rchi)}_{\rm F}) = {\cal N}(k^{(\rchi)}_{\rm F}/b)
\nonumber\\
&\times 
\begin{pmatrix}
1+3\cos(2\phi) & 0 & 3\sin(2\phi)
\\
0&-2&0
\\
3\sin(2\phi) &0& 1-3\cos(2\phi)
\end{pmatrix} 
\end{align} 
where ${\cal N}(x) = [(1-x  )^3 \Theta(x-1)\Theta(2-x) - 1]/6$ and ${\cal N}(0) = -1/6$. 
The explicit form of ${\cal N}(x)$ implies that doping-induced corrections appear only in the limit $k^{(\rchi)}_{\rm F}>b$. Since usually $k^{(\rchi)}_{\rm F}\ll b$, we can safely evaluate $\bm{\mathcal M}_{\rm cv}({\bm b},k^{(\rchi)}_{\rm F})$ for $k^{(\rchi)}_{\rm F}/b = 0$. 

We now proceed to evaluate charge and node anomalous photocurrents for the discrete distributions of Weyl-Berry monopoles illustrated in Fig.~\ref{fig:wsm}. Given such a distribution of $N$ monopoles located at positions ${\bm k} = {\bm b}_{i}$ with effective charge $\rchi_{i}$, we can define the Weyl-Berry dipole and quadrupole moments:
\begin{align}\label{eq:WB_dipole_quadrupole}
&D_\alpha = \sum_{i=1}^{N} \rchi_{i} b_{i,\alpha}~, \nonumber\\
&Q_{\alpha\beta} = \sum_{i=1}^{N} \rchi_i (3 b_{i,\alpha}b_{i,\beta} - b^2_i \delta_{\alpha\beta})~,
\end{align} 
where $b_{i,\alpha}$ denotes the $\alpha$-th Cartesian component of the vector ${\bm b}_i$ and $b^2_i \equiv \sum_{\alpha} b^2_{i,\alpha}$. We note that, upon shifting the origin of momentum space by a vector ${\bm K}$, i.e.~by shifting ${\bm b}_i\to {\bm b}_i+{\bm K}$, the dipole and quadrupole moments undergo the following changes: ${\bm D} \to {\bm D} +C {\bm K}$ and 
$Q_{\alpha\beta} \to Q_{\alpha\beta} + 3(D_\alpha K_\beta+D_\beta K_\alpha)-2{\bm D}\cdot{\bm K}\delta_{\alpha\beta}+C (3 K_{\alpha}K_{\beta} - K^2 \delta_{\alpha\beta})$. Here, $C=\sum_i \rchi_i $ is the total chiral charge, which, as stated above, is guaranteed to be zero in a WSM. For the pure quadrupole in Fig.~\ref{fig:wsm}(b), also the associated total dipole ${\bm D}$ vanishes. We therefore conclude that the pure dipole and quadrupole moments in Fig.~\ref{fig:wsm} are invariant under shifts of the origin.

In an undoped WSM with broken time-reversal symmetry---Fig.~\ref{fig:wsm}(a)---the second-order charge conductivity vanishes after summing over the two Weyl nodes, i.e.~${\bm \sigma}^{(2),{\rm c}}(\omega) = \sum^{2}_{i=1} {\bm \sigma}^{(2)}_{(\rchi_i,{\bm b}_i)}(\omega)= 0$ for ${\bm b}_1= -{\bm b}_2=b\hat{\bm k}_x$ and $\rchi_1=-\rchi_2=\rchi$. However, the second-order node conductivity is finite and given by ${\bm \sigma}^{(2),{\rm n}}(\omega) = \sum^{2}_{i=1}  (\rchi_{i}/e) {\bm \sigma}^{(2)}_{(\rchi_i,{\bm b}_i)}(\omega) = [e^2/ (\omega h^2)]{\rm diag}[0,1,1]$. This implies that the second-order node current in WSMs with broken time-reversal symmetry is along ${\bm{\mathcal F}}(\omega)$ and perpendicular to the dipole vector ${\bm D}$ in Eq.~(\ref{eq:WB_dipole_quadrupole}),
\begin{align}
\sigma^{(2),{\rm n}}_{\alpha\beta}(\omega) = -  \frac{e^2}{h^2 \omega }  \hat{\bm k}_{\alpha}\cdot \left ( \hat{\bm D} \times [\hat{\bm D}\times \hat {\bm k}_\beta ] \right )~.
\end{align}
Notice that the laboratory, position-space, and momentum-space frames have been set to be the same. 

For an undoped WSM with broken inversion symmetry described by the Weyl-Berry quadrupole in Fig.~\ref{fig:wsm}(b) we find
\begin{equation}\label{eq:jec_quadrupole}
{\bm \sigma}^{(2),{\rm c}}(\omega)=\sum^{4}_{i=1}  {\bm \sigma}^{(2)}_{(\rchi_i,{\bm b}_i)}(\omega) = -  \frac{2e^3}{ h^2 \omega }  \frac{\bm Q}{d^2}~, 
\end{equation}
where $d=2\sqrt{\smash[b]{b^2_x+b^2_y}}$ is the diagonal of the rectangle in Fig.~\ref{fig:wsm}(b). 
The explicit form of the quadrupole tensor (\ref{eq:WB_dipole_quadrupole}) reads as ${\bm Q}=  \rchi  S \hat{\bm Q}$,where $\hat{\bm Q}= \sum_{ij}\hat{Q} _{ij} \hat{\bm k}_i \hat{\bm k}_j$ is a dyadic tensor with $\hat{Q}_{xz}=\hat{Q}_{zx}=1$ and $\hat{Q}_{ij}=0$ otherwise. Here, $S=4b_{x} b_{z}$ is the rectangle area and $\rchi$ is the chirality of the node located at ${\bm b}= b_{x} \hat{\bm k}_x - b_{z} \hat{\bm k}_z$ measured from the rectangle center. Since all the diagonal elements of the quadrupole tensor ${\bm Q}$ are zero, the anomalous  charge photocurrent is transverse to ${\bm{\mathcal F}}(\omega)$. Finally, note that when $b_{x}$ or $b_{z}$ vanish, Weyl nodes morph into doubly-degenerate Dirac crossings~\cite{Yan_2017}. In this case, ${\bm Q}$ and ${\bm \sigma}^{(2),{\rm c}}(\omega)$ vanish. 
The second-order node conductivity is also finite and given by
\begin{align}\label{eq:jnc_quadrupole}
{\bm \sigma}^{(2),{\rm n}}(\omega)  
=   \frac{e^2}{h^2\omega} \left \{ \mathbb{1} + \hat{\bm n}\hat{\bm n} - g\hat{\bm n} \times \hat{\bm Q} \right \}~,
\end{align} 
where $g=(b^2_x-b^2_z)/(b^2_x+b^2_z)$ is a measure of the momentum-space anisotropy, $\hat{\bm n}= \hat{\bm k}_y $ is a unit vector perpendicular to the plane of the rectangle in Fig.~\ref{fig:wsm}(b).
Using dyadic algebra, it is possible to show that both  $\hat{\bm n} \hat{\bm n}$ and $\hat{\bm n} \times \hat{\bm Q}$ \cite{footnote_dyadic}   are diagonal tensors and therefore the second-order anomalous node photocurrent in WSMs with broken inversion symmetry is longitudinal. 
\begin{figure}[t]
\centering
\includegraphics[width=85mm]{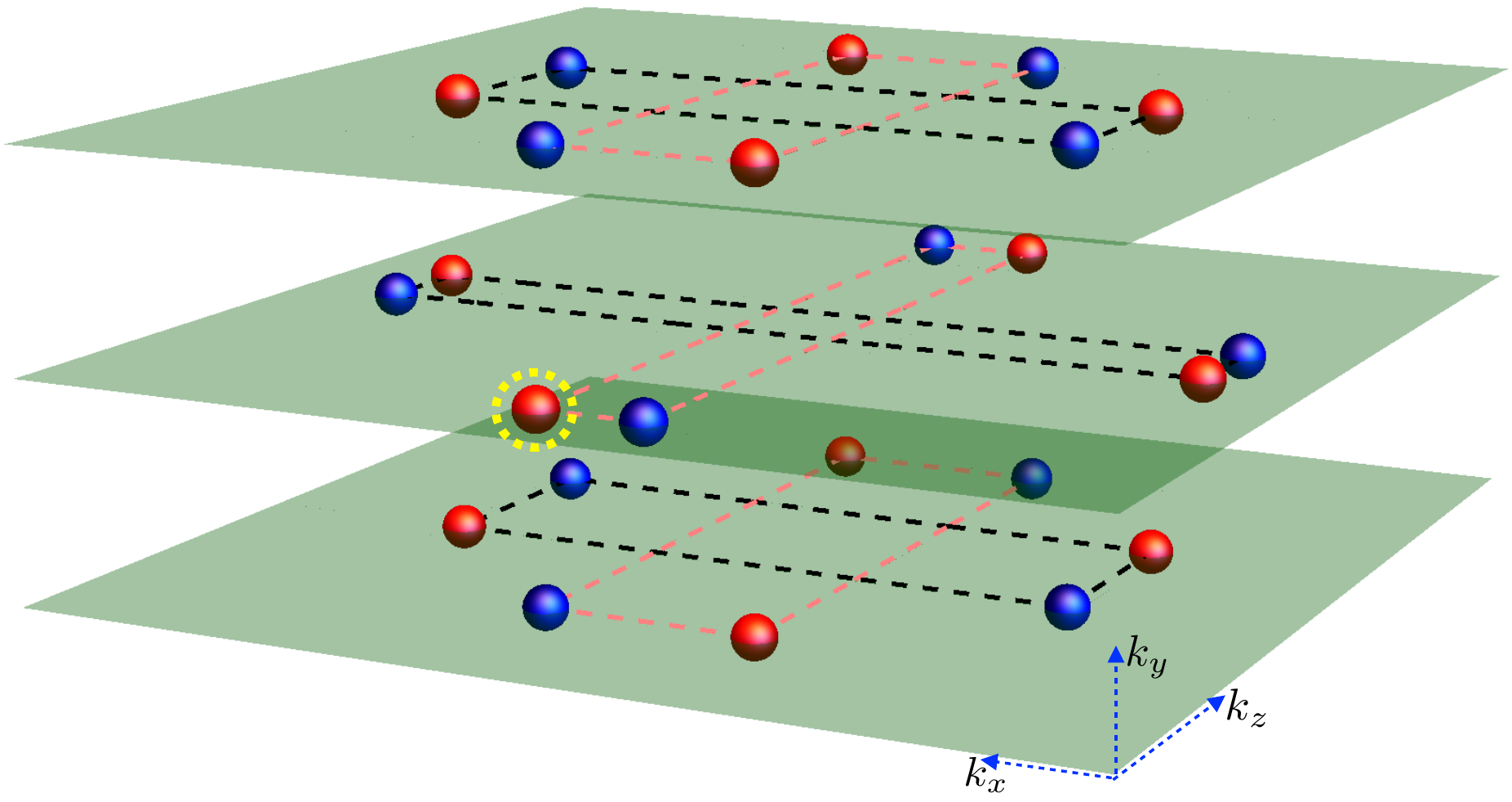}
\caption{A cartoon representing the $N=24$ nodes of TaAs-based WSMs, as distributed in three parallel crystal planes in the BZ~\cite{Yang_np_2015,Xu_np_2015}. Red and blue colors indicate opposite chiral charge, $\chi=\pm$. 
The node located at ${\bm b}= b_{0,x} \hat{\bm k}_x - b_{0,z} \hat{\bm k}_z$, whose chirality is $\rchi$, is specified by a dashed yellow circle.\label{fig:TaAs}}
\end{figure}
\section{Discussion and conclusion}
In terms of actual materials, few WSMs with broken time-reversal symmetry have been studied experimentally. The authors of Ref.~[\onlinecite{Shekhar_arXiv_2016}] have studied 
magnetic lanthanide half-Heusler compounds. However, in these materials, also inversion symmetry is broken and therefore the description of their physical properties goes beyond the theory developed here. 

Time-reversal-invariant WSMs with broken inversion symmetry include the TaAs family with tetragonal lattice symmetry~\cite{Yan_2017}. 
Their cubic BZ contains three parallel layers of Weyl nodes, for a total of $N=24$ nodes---see Fig.~\ref{fig:TaAs}. We set  $\hat{\bm n} = \hat{\bm k}_y$ as the unit vector perpendicular to these planes. This monopole distribution respects two mirror, one fourfold rotational symmetries and it is time-reversal invariant.
%This monopole distribution respects two mirror symmetries, it is four-fold rotational and time-reversal invariant. 
Each of these three layers contains $N=8$ nodes, which can be decomposed into two independent quadrupoles with opposite orientation, $\pm{\bm Q}_{i}$, for consistency with the four-fold rotational symmetry. We use $i=-1,0,+1$ to label bottom, middle, and top planes, respectively.
We therefore immediately see that the total Weyl-Berry quadrupole vanishes in pristine TaAs, implying no anomalous charge photocurrent. For the case of the node photocurrent, however, our theory predicts a second-order anomalous node conductivity
\begin{equation}\label{eq:universal}
{\sigma}^{(2),{\rm n}}_{\alpha\beta}(\omega) = \frac{ N e^2}{4 h^2\omega} (\delta_{\alpha\beta}+\hat{n}_{\alpha}\hat{n}_{\beta})~,
\end{equation}
where $N=24$ is the number of nodes in the TaAs family of WSMs. This is another important result of this work. Note that the first-order anomalous node current vanishes in WSMs with either broken time-reversal or broken inversion symmetry (see Appendix~\ref{app:linear}).  
In this regard, the second-order anomalous node photocurrent is the leading term.
Similar techniques to those used in Ref.~[\onlinecite{Ma_arxiv_2017}] to measure the injection-current-rate contribution to the charge photocurrent can be used to probe the node photocurrent and test the validity of Eq.~(\ref{eq:universal}).

Breaking the four-fold rotational symmetry of TaAs by applying a uniaxial strain 
${\bm \varepsilon}={\rm diag}[\varepsilon,0,0]$ perpendicular to $\hat{\bm n}$ and neglecting Poisson's ratio, we find a finite anomalous second-order charge conductivity given by 
(see Appendix~\ref{app:TaAs})  
\begin{align}
{\bm \sigma}^{(2),{\rm c}}(\omega)  \approx - \varepsilon \frac{4   e^3}{ h^2 \omega }  \left(g_0 \frac{{\bm Q}_0}{d^2_0}- \sum_{i=\pm 1} g_{i} \frac{{\bm Q}_{i}}{d^2_i}\right)~,
\end{align}
where $d_{i}$ and $g_{i}$ are the equivalent of the single-quadrupole quantities $d$ and $g$ defined earlier for ${\bm Q}_{i}$. For TaAs~\cite{Yang_np_2015}, we typically have $(b_{0,x},b_{0,z})\sim(0.10,1.85){\rm \AA}^{-1}$  and $(b_{\pm,x},b_{\pm,z})\sim(0.20,1.15){\rm \AA}^{-1}$. Our theory therefore predicts a non-vanishing anomalous charge conductivity given by $\sigma^{(2),{\rm c}}_{xz}(\omega) =\sigma^{(2),{\rm c}}_{zx}(\omega)  \sim - \rchi \varepsilon e^3/(h^2\omega)$ where $\rchi$ is the chirality of the node in the middle plane located at ${\bm b}= b_{0,x} \hat{\bm k}_x - b_{0,z} \hat{\bm k}_z$. 
Considering $\tau\sim 1{\rm ps}$ as a typical scattering time scale in a clean sample at low temperature, from single-particle scattering mechanisms induced by disorder, our theory can be used for $\hbar\omega>\hbar/\tau \sim 1{\rm meV}$. Since the anomalous nonlinear conductivity scales as $1/\omega$, it is better to use a low-frequency laser in order to achieve a stronger anomalous photocurrent signal. Following Ref.~[\onlinecite{Ma_arxiv_2017}] as an experimental work, one may consider a mid-infrared scanning photocurrent spectroscopy using a CO$_2$ laser with frequency $\hbar\omega_0\sim 120~$meV as the probe. Therefore, for 3\% of strain, we find $|\sigma^{(2),{\rm c}}_{xz}(\omega_0)|\sim 10^{13}~e{\rm V}^{-2} {\rm s}^{-1}$.
\section*{Acknowledgment}
This work was supported by Fondazione Istituto Italiano di Tecnologia.
\appendix
\newpage
\begin{widetext}
\section{The electronic polarization operator} \label{app:pol}
We consider a system of independent spinless fermions in a crystal defined by a generic Hamiltonian ${\cal H}_{0}$.
We introduce Bloch states and band energies,
${\cal H}_0 \phi_{n}({\bm k},{\bm r}) = E_{n}({\bm k})\phi_{n}({\bm k},{\bm r})$,
where ${\bm k}$ is the crystal wavevector and $n$ is a band index. As usual, each Bloch state can be written as
$
\phi_{n}({\bm k},{\bm r})  = \left \langle {\bm r} | n {\bm k}\right\rangle=  u_{n {\bm k}}({\bm r})  e^{i {\bm k}\cdot {\bm r}}$,
where $u_{n {\bm k}}({\bm r})$ is a periodic function of ${\bm r}$. We assume that the system is subject to an external homogeneous and time dependent electric field $\bm{\mathcal E}(t)$, which we describe in the length gauge~\cite{Aversa_prb_1995}, i.e.~by using a scalar potential of the form $V({\bm r},t)=- e {\bm r} \cdot \bm{\mathcal E}(t)$. 
Therefore, the second quantized Hamiltonian reads as following,
\begin{equation}\label{eq:H_psi}
\hat {\cal H}(t) = \int d^D{\bm r}~\hat \psi^\dagger({\bm r}) \left [ {\cal H}_0 - e {\bm r} \cdot \bm{\mathcal E}(t)\right ] \hat \psi({\bm r})~, 
\end{equation}
where $D$ is the dimensionality of the system and the field operator $\hat{\psi}({\bm r})$ can be written in terms of Bloch-state annihilation operators, $\hat a_{n}({\bm k})$, in the usual manner:
\begin{equation}\label{eq:psi_to_a}
\hat \psi({\bm r})  = \sum_{n, {\bm k}} \hat a_{n}({\bm k}) \phi_{n}({\bm k},{\bm r})~.
\end{equation}
Here,  
\begin{equation}
\sum_{\bm k} \equiv \int_{\rm BZ} \frac{d^{D}{\bm k}}{(2\pi)^{D}}~,
\end{equation}
where ``${\rm BZ}$'' is a shorthand for ``Brillouin zone''. We introduce
\begin{equation}\label{eq:scalar_product}
 \left \langle n {\bm k} | m {\bm k}'\right\rangle  \equiv \int d^{D}{\bm r}~\phi^\ast_n({\bm k},{\bm r}) \phi_m({\bm k}',{\bm r})  = \delta_{nm } \delta({\bm k} - {\bm k}')~,
\end{equation}
\begin{equation}
\left \langle n {\bm k} |{\bm r} |m {\bm k}'\right\rangle \equiv  \int d^{D}{\bm r}~\phi^\ast_n({\bm k},{\bm r}) {\bm r} \phi_m({\bm k}',{\bm r})~,
\end{equation}
and the following useful identity
\begin{equation}
i  \frac{\partial \phi_{n}({\bm k}, {\bm r})}{\partial {\bm k}}  = i  e^{i {\bm k}\cdot {\bm r} } \frac{\partial u_{n {\bm k}}({\bm r})}{\partial {\bm k}}    - {\bm r} \phi_{n}({\bm k},{\bm r})~.
\end{equation}
Also, we define the quantity $\bm{\mathcal A}_{mn}({\bm k})$ as following
\begin{equation}
i   \frac{\partial u_{n {\bm k}}({\bm r})}{\partial {\bm k}}   \equiv \sum_{m}  u_{m {\bm k}}({\bm r})\bm{\mathcal A}_{mn}({\bm k})~.
\end{equation}
We therefore have
\begin{equation}
 i  \frac{\partial \phi_{n}({\bm k}, {\bm r}) }{\partial {\bm k}} = \sum_{m} \phi_{m}( {\bm k}, {\bm r})  \bm{\mathcal A}_{mn}({\bm k})   - {\bm r} \phi_{n}({\bm k},{\bm r})~.
\end{equation}
Multiplying both members of the previous equation by $\phi^\ast_{n'}({\bm k}', {\bm r})$ and integrating over ${\bm r}$ we find:
\begin{equation}
 i \int d^{D}{\bm r} \phi^\ast_{n'}({\bm k}', {\bm r})  \frac{\partial \phi_{n}({\bm k}, {\bm r})}{\partial {\bm k}}  =
  \sum_{m}   \int d^{D} {\bm r}  \phi^\ast_{n'}({\bm k}', {\bm r})  \phi_{m}( {\bm k}, {\bm r})  \bm{\mathcal A}_{mn}({\bm k})   
  -   \int d^{D} {\bm r}  \phi^\ast_{n'}({\bm k}', {\bm r})  {\bm r} \phi_{n}({\bm k},{\bm r})~.
\end{equation}
Using Eq.~(\ref{eq:scalar_product}) we therefore conclude that 
\begin{equation}\label{eq:r_element}
\left \langle n' {\bm k}' |{\bm r} |n {\bm k} \right\rangle = \bm{\mathcal A}_{n'n}({\bm k})  \delta({\bm k}-{\bm k}') -  i \delta_{n'n} \frac{\partial }{\partial {\bm k}}  \delta({\bm k}-{\bm k}')~,
\end{equation}
with $\bm{\mathcal A}_{n'n}({\bm k})  = i \left \langle u_{n' {\bm k}} |  {\partial u_{n {\bm k}} }/{\partial {\bm k}} \right \rangle$ and $\bm{\mathcal A}_{nn}({\bm k})$ coinciding with the usual Berry connection~\cite{Sundaram_Niu,Xiao_rmp_2010}. 
By inserting Eq.~(\ref{eq:psi_to_a}) into Eq.~(\ref{eq:H_psi}) and using Eq.~(\ref{eq:r_element}), we arrive at the following Hamiltonian 
\begin{eqnarray}
\hat {\cal H}(t) &=&    \sum_{n, {\bm k}} E_{n}({\bm k})\hat a^\dagger_{n}({\bm k}) \hat a_{n}({\bm k})
-e \bm{\mathcal E}(t) \cdot  \sum_{n, m, {\bm k}}  \bm{\mathcal A}_{nm}({\bm k}) \hat a^\dagger_{n}({\bm k}) \hat a_{m}({\bm k}) \nonumber\\
&+& i e \bm{\mathcal E}(t) \cdot  \sum_{n, {\bm k}, {\bm k'}} \hat a^\dagger_{n}({\bm k}) \hat a_{n}({\bm k}') \frac{\partial }{\partial {\bm k}'}  \delta({\bm k}-{\bm k}')~.
\end{eqnarray}
After performing an integration by parts in the last term of the previous equation, we find 
\begin{eqnarray}
 \hat {\cal H}(t) =    \sum_{n,{\bm k}} E_{n}({\bm k})\hat a^\dagger_{n}({\bm k}) \hat a_{n}({\bm k})
-e \bm{\mathcal E}(t) \cdot  \sum_{n, m, {\bm k}} \bm{\mathcal A}_{nm}({\bm k}) \hat a^\dagger_{n}({\bm k}) \hat a_{m}({\bm k}) 
- i e \bm{\mathcal E}(t) \cdot  \sum_{n, {\bm k}}  \hat a^\dagger_{n}({\bm k}) \frac{\partial }{\partial {\bm k}}  \hat a_{n}({\bm k})~.  
\end{eqnarray}
We can therefore rewrite the Hamiltonian in the following compact form,
$ \hat {\cal H}(t)= \hat {\cal H}_0 - \hat {\bm P} \cdot \bm{\mathcal E}(t)$,
where the electric polarization operator is given by  
\begin{equation}
 \hat {\bm P}=e \sum_{n, {\bm k}} \hat a^\dagger_{n}({\bm k})
 \left [i\frac{\partial  \hat a_{n}({\bm k}) }{\partial {\bm k}}+ \sum_{m} \bm{\mathcal A}_{nm}({\bm k}) \hat a_{m}({\bm k})\right ]~.
\end{equation}
\section{Non-perturbative expression for the intra- and inter-band contributions to the charge current}\label{app:current}
From the time dependence of the macroscopic polarization ${\bm P}(t) = \langle  \hat {\bm P} \rangle$, one can evaluate the macroscopic charge current:
\begin{equation}\label{eq:J_dpdt_appendix}
{\bm J}(t)= \frac{\partial {\bm P}(t) }{\partial t} = e  \sum_{ {\bm k}}  \left \{ 
\sum_{n} \left [\bm{\mathcal A}_{nn}({\bm k}) \frac{\partial \rho_{nn}({\bm k})}{\partial t} + \frac{\partial {\bm \zeta}_{nn}({\bm k})}{\partial t} \right ]
+ \sum_{m\neq n}  \bm{\mathcal A}_{nm}({\bm k}) \frac{\partial \rho_{nm}({\bm k})}{\partial t} 
 \right \}~,
\end{equation}
where 
$\rho_{nm}({\bm k}) = \left \langle \hat a^\dagger_{n}({\bm k}) \hat a_{m} ({\bm k})\right \rangle$
and
${\bm \zeta}_{nm}({\bm k}) =  i \left \langle \hat a^\dagger_{n}({\bm k})  {\partial \hat a_{m} ({\bm k}) }/{\partial {\bm k}} \right \rangle$.

We now need to derive equations of motion for the density matrix $\rho_{nm}({\bm k})$ and the quantity ${\bm \zeta}_{nm}({\bm k})$. To this end, we utilize Heisenberg's equation of motion,  
$i\hbar  {\partial \hat A}/{ \partial t} =   [ \hat A , \hat{\cal H}   ]$.
After straightforward calculations, we obtain the following identities:
\begin{eqnarray}
&&  [ \hat a_{m}({\bm k}) , \hat{\cal H}] =  
 E_{m}({\bm k})  \hat a_{m}({\bm k})   
-e\bm{\mathcal E}(t) \cdot  \sum_{p}  {\cal\bm A}_{m p}({\bm k}) \hat a_{p}({\bm k}) 
-  i e\bm{\mathcal E}(t) \cdot   \frac{\partial }{\partial {\bm k}}  \hat a_{m}({\bm k})~,  
\\
&&  [ \hat a^\dagger_{n}({\bm k}) , \hat{\cal H} ]=
-   E_{n}({\bm k})  \hat a^\dagger_{n}({\bm k}) 
+e \bm{\mathcal E}(t) \cdot  \sum_{p} {\cal\bm A}_{pn}({\bm k})   \hat a^\dagger_{p}({\bm k})   
-  i e \bm{\mathcal E}(t) \cdot  \frac{\partial }{\partial {\bm k}}  \hat a^\dagger_{n}({\bm k})~.
 \end{eqnarray}
Using these relations, we find
\begin{eqnarray} \label{eq:rho_nn}
 { \frac{\partial \rho_{nn}({\bm k})}{\partial t}=
 -  \frac{e \bm{\mathcal E}(t)}{\hbar} \cdot  \frac{\partial \rho_{nn}({\bm k}) }{\partial {\bm k}} 
- i \frac{e\bm{\mathcal E}(t)}{\hbar} \cdot  \sum_{m (\neq n)}  
 [\bm{\mathcal A}_{mn}({\bm k})   \rho_{mn}({\bm k})  - \bm{\mathcal A}_{n m}({\bm k}) \rho_{n m}({\bm k})] }
 \end{eqnarray}
and
\begin{eqnarray} 
 \frac{\partial {\bm \zeta}_{nn}({\bm k})}{\partial t} &=&
\frac{1}{\hbar} \rho_{nn}({\bm k}) \frac{\partial }{\partial {\bm k}} \left [E_{n}({\bm k}) - e\bm{\mathcal E}(t) \cdot \bm{\mathcal A}_{nn}({\bm k}) \right ] 
- \frac{e}{\hbar} \sum_{m (\neq n)}   \rho_{nm}({\bm k}) \frac{\partial }{\partial {\bm k}} [\bm{\mathcal E}(t) \cdot  \bm{\mathcal A}_{n m}({\bm k})] 
\nonumber \\
&-& \frac{e}{\hbar}  \bm{\mathcal E}(t) \cdot    \frac{\partial }{\partial {\bm k}}    {\bm \zeta}_{nn}({\bm k})   
- i \frac{e}{\hbar} \bm{\mathcal E}(t) \cdot \sum_{m (\neq n)} [ \bm{\mathcal A}_{mn}({\bm k}) {\bm \zeta}_{mn}({\bm k})  -   \bm{\mathcal A}_{nm}({\bm k}) 
{\bm \zeta}_{nm}({\bm k})]~.
 \end{eqnarray}
Using the above results, we find 
\begin{eqnarray}\label{eq:intra}
\sum_{n} \left [\bm{\mathcal A}_{nn}({\bm k}) \frac{\partial \rho_{nn}({\bm k})}{\partial t} + \frac{\partial {\bm \zeta}_{nn}({\bm k})}{\partial t} \right ] 
&=&   -  \sum_{n}\bm{\mathcal A}_{nn}({\bm k})  \frac{e \bm{\mathcal E}(t)}{\hbar} \cdot  \frac{\partial \rho_{nn}({\bm k})}{\partial {\bm k}} 
\nonumber \\
&-& i  \sum_{n} \bm{\mathcal A}_{nn}({\bm k}) \frac{e \bm{\mathcal E}(t)}{\hbar} \cdot  \sum_{m(\neq n)} [\bm{\mathcal A}_{mn}({\bm k})   \rho_{mn}({\bm k})  - \bm{\mathcal A}_{n m}({\bm k}) \rho_{n m}({\bm k})]
 \nonumber \\
 &+&
 \sum_{n} \frac{1}{\hbar}\rho_{nn}({\bm k}) \frac{\partial }{\partial {\bm k}} \left [E_{n}({\bm k}) - e \bm{\mathcal E}(t) \cdot \bm{\mathcal A}_{nn}({\bm k}) \right ] 
\nonumber \\
&-&  \sum_{n}\frac{e}{\hbar} \sum_{m(\neq n)}  \rho_{nm}({\bm k}) \frac{\partial }{\partial {\bm k}}  [\bm{\mathcal E}(t) \cdot   {\cal\bm A}_{n m}({\bm k})] 
\nonumber \\
&-&  \sum_{n}\frac{e}{\hbar}  \bm{\mathcal E}(t) \cdot    \frac{\partial {\bm \zeta}_{nn}({\bm k})   }{\partial {\bm k}}
\nonumber \\
&-& i  \sum_{n}\frac{e}{\hbar} \bm{\mathcal E}(t) \cdot \sum_{m(\neq n)}[\bm{\mathcal  A}_{mn}({\bm k}) {\bm \zeta}_{mn}({\bm k})  -   \bm{\mathcal A}_{nm}({\bm k}) 
{\bm \zeta}_{nm}({\bm k})]~.
\end{eqnarray}
We also note the following useful identity:
\begin{equation}\label{eq:berry}
\frac{\partial }{\partial {\bm k}} [\bm{\mathcal E}(t) \cdot \bm{\mathcal A}_{nn}({\bm k})] = 
\bm{\mathcal E}(t) \cdot \frac{\partial }{\partial {\bm k}} \bm{\mathcal A}_{nn}({\bm k}) 
+ 
\bm{\mathcal E}(t) \times {\bm \Omega}_{n}({\bm k})~,
\end{equation}
where ${\bm \Omega}_{n}({\bm k})=  {\partial }/{\partial {\bm k}} \times \bm{\mathcal A}_{nn}({\bm k}) $ is the Berry curvature.  
Inserting Eq.~(\ref{eq:berry}) in Eq.~(\ref{eq:intra}), we find 
\begin{eqnarray}
\sum_{n} \left [\bm{\mathcal A}_{nn}({\bm k}) \frac{\partial \rho_{nn}({\bm k})}{\partial t} + \frac{\partial {\bm \zeta}_{nn}({\bm k})}{\partial t} \right ] 
&=&   -  \sum_{n}\bm{\mathcal A}_{nn}({\bm k})  \frac{e \bm{\mathcal E}(t)}{\hbar} \cdot  \frac{\partial \rho_{nn}({\bm k})}{\partial {\bm k}}    
\nonumber \\
&-& i  \sum_{n} \bm{\mathcal A}_{nn}({\bm k}) \frac{e \bm{\mathcal E}(t)}{\hbar} \cdot  \sum_{m (\neq n)} [\bm{\mathcal A}_{mn}({\bm k})   \rho_{mn}({\bm k})  - \bm{\mathcal A}_{n m}({\bm k}) \rho_{n m}({\bm k})] 
 \nonumber \\
 &+&
 \sum_{n} \frac{1}{\hbar}\frac{\partial E_{n}({\bm k})  }{\partial {\bm k}}  \rho_{nn}({\bm k}) 
-\frac{e}{\hbar}  \sum_{n} [\bm{\mathcal E}(t) \times {\bm \Omega}_{n}({\bm k})] \rho_{nn}({\bm k}) 
-\frac{e}{\hbar}  \sum_{n}\rho_{nn}({\bm k}) \bm{\mathcal E}(t) \cdot \frac{\partial }{\partial {\bm k}} \bm{\mathcal A}_{nn}({\bm k}) 
\nonumber \\ 
&-& \frac{e}{\hbar} \sum_{m (\neq n)}  \sum_{n} \rho_{nm}({\bm k}) \frac{\partial }{\partial {\bm k}} [\bm{\mathcal E}(t) \cdot   \bm{\mathcal A}_{n m}({\bm k})]  
\nonumber \\
&-&  \sum_{n}\frac{e}{\hbar}  \bm{\mathcal E}(t) \cdot    \frac{\partial {\bm \zeta}_{nn}({\bm k})  }{\partial {\bm k}}     
\nonumber \\
&-& i  \frac{e}{\hbar} \bm{\mathcal E}(t) \cdot \sum_{m (\neq n)}  \sum_{n} [\bm{\mathcal A}_{mn}({\bm k}) {\bm \zeta}_{mn}({\bm k})  -   \bm{\mathcal A}_{nm}({\bm k}) 
{\bm \zeta}_{nm}({\bm k})]~.
\end{eqnarray}
At this stage, it is useful to introduce the quantity  $\bm{\mathcal R}_{n}({\bm k})= \bm{\mathcal A}_{nn}({\bm k})\rho_{nn}({\bm k})+  {\bm \zeta}_{nn}({\bm k})$. We find 
\begin{eqnarray}\label{eq:practically_final}
\sum_{n} \left [\bm{\mathcal A}_{nn}({\bm k}) \frac{\partial \rho_{nn}({\bm k})}{\partial t} + \frac{\partial {\bm \zeta}_{nn}({\bm k})}{\partial t} \right ] 
&=&   -   \frac{e \bm{\mathcal E}(t)}{\hbar} \cdot  \sum_{n} \frac{\partial \bm{\mathcal R}_{n}({\bm k})}{\partial {\bm k}}   
 \nonumber \\
 &+&
 \sum_{n} \frac{1}{\hbar}\frac{\partial E_{n}({\bm k})  }{\partial {\bm k}}  \rho_{nn}({\bm k}) 
\nonumber \\
&-&\frac{e}{\hbar}  \sum_{n} [\bm{\mathcal E}(t) \times {\bm \Omega}_{n}({\bm k})]\rho_{nn}({\bm k})  
\nonumber \\
&-& i  \sum_{n} \bm{\mathcal A}_{nn}({\bm k}) \frac{e\bm{\mathcal E}(t)}{\hbar} \cdot  \sum_{m (\neq n)}  
[\bm{\mathcal A}_{mn}({\bm k})   \rho_{mn}({\bm k})  - \bm{\mathcal A}_{n m}({\bm k}) \rho_{n m}({\bm k})]
\nonumber \\ 
&-& \frac{e}{\hbar} \sum_{n}\sum_{m (\neq n)}  \rho_{nm}({\bm k}) \frac{\partial }{\partial {\bm k}}  [ \bm{\mathcal E}(t) \cdot   \bm{\mathcal A}_{n m}({\bm k})] 
\nonumber \\
&-& i  \frac{e}{\hbar} \bm{\mathcal E}(t) \cdot \sum_{n} \sum_{m (\neq n)}  [\bm{\mathcal A}_{mn}({\bm k}) {\bm \zeta}_{mn}({\bm k})  -  \bm{\mathcal A}_{nm}({\bm k}) 
{\bm \zeta}_{nm}({\bm k})]~.
\end{eqnarray}
By considering the exchange of dummy band indices, $m \leftrightarrow n$, it is easy to see that the last term on the right-hand side of the previous equation vanishes.
We can now collect together in a compact fashion all the inter-band terms on the right-hand side of Eq.~(\ref{eq:practically_final}) by introducing the following generalized derivative: $\hat{\bm{\mathcal D}}_{mn}({\bm k})=   {\partial }/{\partial {\bm k}} + i[ {\cal\bm A}_{mm}({\bm k})- {\cal\bm A}_{nn}({\bm k})]$. We find
\begin{eqnarray}\label{eq:jnn}
\sum_{n} \left [\bm{\mathcal A}_{nn}({\bm k}) \frac{\partial \rho_{nn}({\bm k})}{\partial t} + \frac{\partial {\bm \zeta}_{nn}({\bm k})}{\partial t} \right ] 
&=&   - \frac{e \bm{\mathcal E}(t) }{\hbar} \cdot \sum_{n} \frac{\partial \bm{\mathcal R}_{n}({\bm k})}{\partial {\bm k}}   
\nonumber \\
 &+&  \frac{1}{\hbar} \sum_{n}\frac{\partial E_{n}({\bm k})  }{\partial {\bm k}}  \rho_{nn}({\bm k}) 
\nonumber \\
&-&\frac{e}{\hbar}  \sum_{n} [\bm{\mathcal E}(t) \times {\bm \Omega}_{n}({\bm k})] \rho_{nn}({\bm k})  
\nonumber \\ 
&-& \frac{e}{\hbar}  \sum_{m \neq n }  \rho_{nm}({\bm k}) \hat{\bm {\mathcal D}}_{mn}({\bm k}) \bm{\mathcal E}(t) \cdot  
\bm{\mathcal A}_{n m}({\bm k})~,
\end{eqnarray}
where we have introduced the shortand $\sum_{m \neq n} \mapsto \sum_{n}  \sum_{m(\neq n)}$.  By considering Eqs.~(\ref{eq:J_dpdt_appendix}) and Eq.~(\ref{eq:jnn}) we immediately find Eqs.~(3) and~(4) in the main text.    
\section{Proof of Eq.~(5)}\label{app:eq5}
Introducing the zeroth-order contribution $\rho^{(0)}_{nm}({\bm k})=\delta_{nm} n_{\rm F}(E_{n}({\bm k}))$ to the density matrix, and the first-order contribution $\rho^{(1)}_{nm}({\bm k})$, we find the first- and second-order anomalous (i.e.~Berry-curvature-controlled) contributions to the current response: 
\begin{equation}
{\bm J}^{(1)}(t) = -\frac{e^2}{\hbar}  \bm{\mathcal E}(t) \times  \sum_{n {\bm k}}  {\bm \Omega}_{n}({\bm k}) n_{\rm F}(E_{n}({\bm k}))
\end{equation}
and
\begin{equation}
{\bm J}^{(2)}(t) = -\frac{e^2}{\hbar}  {\cal \bm E}(t) \times  \sum_{n {\bm k}}  {\bm \Omega}_{n}({\bm k}) \rho^{(1)}_{nn}({\bm k})~.
\end{equation}
In the frequency domain, we have  
\begin{equation}\label{eq:linear-response-frequency-domain}
{\bm J}^{(1)}(\omega) = -\frac{e^2}{\hbar}   \sum_{n {\bm k}}  {\cal \bm E}(\omega) \times {\bm \Omega}_{n}({\bm k}) n_{\rm F}(E_{n}({\bm k}))
\end{equation}
and for a single-frequency driving electric field we end up with
\begin{equation}
{\bm J}^{(2)}(\omega_\Sigma) = -\frac{e^2}{2\hbar}   \sum_{n, {\bm k}}  \left \{ {\cal \bm E}(\omega_1) \times {\bm \Omega}_{n}({\bm k}) \rho^{(1)}_{nn}({\bm k},\omega_2) +  \omega_1 \leftrightarrow \omega_2 \right \}~,
\end{equation}
where $\omega_\Sigma=\omega_1+\omega_2$. 
The first-order contribution to the (intra-band) density matrix can be immediately found from Eq.~(\ref{eq:rho_nn}):
\begin{equation}
\frac{\partial \rho^{(1)}_{nn}({\bm k})}{\partial t} = -  \frac{e \bm{\mathcal E}(t)}{\hbar} \cdot    \frac{\partial \rho^{(0)}_{nn}({\bm k})}{\partial {\bm k}}~,
\end{equation}
or, in the frequency domain, 
\begin{equation} 
\rho^{(1)}_{nn}({\bm k},\omega) =-  \frac{i}{ \omega} \frac{e}{\hbar} \bm{\mathcal E}(\omega) \cdot  \frac{\partial n_{\rm F}(E_{n}({\bm k}))}{\partial {\bm k}}~. 
 \end{equation}  
The anomalous contribution to the second-order current therefore reads 
\begin{equation}
{\bm J}^{(2)}(\omega_\Sigma) = i \frac{e^3}{2\hbar^2}   \sum_{n, {\bm k}}  \left[ \bm{\mathcal E}(\omega_1) \times {\bm \Omega}_{n}({\bm k})  
\frac{\bm{\mathcal E}(\omega_2)}{ \omega_2} \cdot  \frac{\partial n_{\rm F}(E_{n}({\bm k}))}{\partial {\bm k}}   +  \omega_1 \leftrightarrow \omega_2 \right]~.
\end{equation}
Integrating by parts, we finally reach the desired result:
\begin{equation}
{\bm J}^{(2)}(\omega_\Sigma) = - i \frac{e^3}{2\hbar^2}   \sum_{n, {\bm k}} n_{\rm F}(E_{n}({\bm k}))  \left\{ 
\frac{\bm{\mathcal  E}(\omega_1)}{ \omega_2}  \cdot  \frac{\partial}{\partial {\bm k}} \left [\bm {\mathcal E}(\omega_2) \times {\bm \Omega}_{n}({\bm k}) \right ]   +  \omega_1 \leftrightarrow \omega_2 \right\}~.
\end{equation}
Note that the boundary term stemming from the integration by parts vanishes because of the vanishing of the Fermi function at infinity. 
In this work we use a continuum low-energy description and lattice effects related to the presence of a finite-size BZ zone have been neglected.
\section{Proof of Eq.~(6)}\label{app:eq6}
The anomalous second-order contribution to the dc photocurrent was found to be
\begin{equation}\label{eq:second-order-dc-photocurrent}
J^{(2)}_\alpha(0) = \frac{e^3}{\hbar^2 \omega}\int \frac{d^3 {\bm k}}{(2\pi)^3} \sum_{n}
\sum_{ \beta\gamma\delta} \epsilon_{\alpha\gamma \delta} 
 {\rm Im} \left \{  {\cal E}_{\beta}(\omega) {\cal   E}^\ast_{\gamma}(\omega)  \right \}  
\frac{\partial   { \Omega}_{n,\delta}({\bm k}) }{\partial k_\beta}  n_{\rm F}(E_{n}({\bm k})) ~. 
\end{equation}
We decompose the electric field product in symmetric and anti-symmetric parts:
\begin{equation}
 {\cal  E}_\beta(\omega) {\cal E}^\ast_\gamma(\omega) =
 {\rm Re} \left [ {\cal  E}_\beta(\omega) {\cal E}^\ast_\gamma(\omega) \right ] + \frac{1}{2} \sum_{\ell} \epsilon_{\ell \beta \gamma} \{ {\bm {\mathcal  E}}(\omega)\times {\bm {\mathcal  E}}^\ast(\omega)\}_{\ell}~.
 \end{equation}
We therefore find 
\begin{equation}\label{eq:imaginary_part}
{\rm Im} \left \{{\cal  E}_\beta(\omega) {\cal E}^\ast_\gamma(\omega) \right \}=
 -\frac{1}{2} \sum_{\ell} \epsilon_{\ell \beta \gamma} {\cal  \bm F}_{\ell}(\omega) 
 \end{equation}
where ${\bm {\mathcal  F}}(\omega) = i {\bm {\mathcal  E}}(\omega)\times {\bm {\mathcal  E}}^\ast(\omega)$. Using Eq.~(\ref{eq:imaginary_part}) in Eq.~(\ref{eq:second-order-dc-photocurrent}) we find
\begin{equation} \label{eq:double-levi-civita}
 J^{(2)}_\alpha(0) =- \frac{e^3}{2\hbar^2 \omega}\int \frac{d^3 {\bm k}}{(2\pi)^3} 
 \sum_{n}
\sum_{ \beta\gamma\delta \ell } 
~\epsilon_{\alpha\gamma \delta} 
~\epsilon_{\ell \beta \gamma}
~ {\cal  \bm F}_{\ell}(\omega) 
\frac{\partial   { \Omega}_{n,\delta}({\bm k}) }{\partial k_\beta}  n_{\rm F}(E_{n}({\bm k})) ~. 
\end{equation}
Using that
\begin{equation}
- \sum_{\gamma} \epsilon_{\alpha\gamma \delta}  \epsilon_{\ell \beta \gamma} 
= \sum_{\gamma}  \epsilon_{\alpha \delta \gamma}  \epsilon_{\ell \beta \gamma}
= \delta_{\alpha \ell} \delta_{\delta \beta} - \delta_{\alpha \beta} \delta_{\delta \ell}~,
\end{equation}
we find
\begin{equation}
 {\bm J}^{(2)}(0) = \frac{e^3}{2\hbar^2 \omega}\int \frac{d^3 {\bm k}}{(2\pi)^3} \sum_n \left \{
 { \bm {\mathcal F}}(\omega) 
\frac{\partial }{\partial {\bm k} } \cdot { \bm \Omega}_{n}({\bm k})
-
 \frac{\partial }{\partial {\bm k}} [ { \bm {\mathcal F}}(\omega) \cdot {\bm  \Omega}_{n}({\bm k})]
 \right\}  n_{\rm F}(E_{n}({\bm k}))~.
\end{equation}
Finally, exploiting the identity
\begin{equation}
 \frac{\partial }{\partial {\bm k}} [ { \bm {\mathcal F}}(\omega) \cdot {\bm  \Omega}_{n}({\bm k})]
 = 
 { \bm {\mathcal F}}(\omega) \cdot  \frac{\partial {\bm  \Omega}_{n}({\bm k}) }{\partial {\bm k}} 
 + 
 { \bm {\mathcal F}}(\omega) \times  \left(\frac{\partial }{\partial {\bm k}} \times {\bm  \Omega}_{n}({\bm k})\right)~,
\end{equation}
we obtain Eq.~(6) in the main text: 
\begin{equation}
 {\bm J}^{(2)}(0) = \frac{e^3}{2\hbar^2 \omega}\int \frac{d^3 {\bm k}}{(2\pi)^3} 
\sum_n \left [
 { \bm {\mathcal F}}(\omega) 
\frac{\partial }{\partial {\bm k} } \cdot { \bm \Omega}_{n}({\bm k})
-
 { \bm {\mathcal F}}(\omega) \cdot  \frac{\partial {\bm  \Omega}_{n}({\bm k}) }{\partial {\bm k}} 
 - 
 { \bm {\mathcal F}}(\omega) \times \left( \frac{\partial }{\partial {\bm k}} \times {\bm  \Omega}_{n}({\bm k}) \right )
 \right ] n_{\rm F}(E_{n}({\bm k}))~.
\end{equation}
\section{ Proof of Eq.~(11)}\label{app:eq11}
We remind the reader about the definition of $\bm{\mathcal M}_{n=\pm}({\bm b}, k_{\rm F})$ at zero temperature given in the main text:
\begin{equation}
{\mathcal M}_{n, \alpha\beta}({\bm b}, k_{\rm F})= \frac{1}{4\pi} \int d^3{\bm k} \frac{3 ({\bm k}-{\bm b})_\alpha ({\bm k}-{\bm b})_\beta -|{\bm k}-{\bm b}|^2 \delta_{\alpha\beta}}{|{\bm k}-{\bm b}|^5} \Theta(n k_{\rm F}-n |{\bm k}-{\bm b}|)~.
\end{equation}
Let us choose ${\bm b}_0=b \hat{\bm x}$ with $b\neq 0$. These are the non-vanishing elements of the tensor for a fully-filled valence band: 
\begin{equation}
{\mathcal M}_{-,xx}({\bm b}_0,0) 
=  \frac{1}{4\pi}  \int^{\infty}_{0} k^2 dk \int^{1}_{-1} d\cos\theta \int^{2\pi}_{0} d\phi \frac{3 ( k \cos\theta- b)^2 - (k^2+b^2-2k b \cos\theta)  }{(k^2+b^2-2k b \cos\theta)^{\frac{5}{2}}}~~,
\end{equation}
\begin{equation}
{\mathcal M}_{-,yy}({\bm b}_0,0) 
=  \frac{1}{4\pi}  \int^{\infty}_{0} k^2 dk \int^{1}_{-1} d\cos\theta \int^{2\pi}_{0} d\phi \frac{3 ( k \sin\theta\sin\phi)^2 - (k^2+b^2-2k b \cos\theta)  }{(k^2+b^2-2k b \cos\theta)^{\frac{5}{2}}}~~,
\end{equation}
and
\begin{equation}
{\mathcal M}_{-,zz}({\bm b}_0,0) 
= \frac{1}{4\pi}  \int^{\infty}_{0} k^2 dk \int^{1}_{-1} d\cos\theta \int^{2\pi}_{0} d\phi \frac{3 ( k \sin\theta\cos\phi)^2 - (k^2+b^2-2k b \cos\theta) }{(k^2+b^2-2k b \cos\theta)^{\frac{5}{2}}}~.
\end{equation}
The above integrals can be performed analytically:   
\begin{eqnarray}
{\mathcal M}_{-,xx}({\bm b}_0,0) &=& \frac{1}{2} \int^{\infty}_{0} k^2 dk \int^{1}_{-1} d x  \frac{3 ( k  x- b)^2 - (k^2+b^2-2k b x) }{(k^2+b^2-2k b x)^{\frac{5}{2}}}
\nonumber \\ 
&=& \frac{1}{b^3} \int^{\infty}_{0}  dk  k^2 (1+{\rm sign}[b-k]) =  \frac{2}{b^3} \int^{b}_{0} dk k^2  =  \frac{2}{3}
\end{eqnarray}
\begin{eqnarray}
{\mathcal M}_{-,yy}({\bm b}_0,0) ={\cal M}_{-,zz}({\bm b}_0,0)  &=& \frac{1}{2} \int^{\infty}_{0} k^2 dk \int^{1}_{-1} d x   \frac{3/2  k^2 (1-x^2) - (k^2+b^2-2k b x)  }{(k^2+b^2-2k b x)^{\frac{5}{2}}}
\nonumber \\     
&=&-\frac{1}{2b^3} \int^{\infty}_{0} k^2 dk  (1+{\rm sign}[b-k]) = -\frac{1}{b^3} \int^{b}_{0} k^2 dk  = - \frac{1}{3}~.
\end{eqnarray}
In summary, 
\begin{equation}\label{eq:Mv}
{\mathcal M}_{-,xx}({\bm b}_0,0) = \frac{2}{3},~~{\cal M}_{-,yy}({\bm b}_0,0) ={\mathcal M}_{-,zz}({\bm b}_0,0)= -\frac{1}{3},~~ {\mathcal M}_{-,\alpha\neq \beta}({\bm b}_0,0) =0~.
\end{equation}
The contribution due to a partially-filled conduction band is: 
\begin{eqnarray}
{\mathcal M}_{+,xx}({\bm b}_0,k_{\rm F}) &=& \frac{1}{2} \int^{\infty}_{0} k^2 dk \int^{1}_{-1} d x  \frac{3 ( k  x- b)^2 - (k^2+b^2-2k b x) }{(k^2+b^2-2k b x)^{\frac{5}{2}}}   \Theta (k_{\rm F} - \sqrt{k^2+b^2-2k b x})
\nonumber \\ 
&=& \frac{1}{2} \int^{\infty}_{0} k^2 dk \int^{1}_{x_0} d x  \frac{3 ( k  x- b)^2 - (k^2+b^2-2k b x) }{(k^2+b^2-2k b x)^{\frac{5}{2}}}   \Theta (1-x_0) \Theta(x_0+1)~,
\end{eqnarray}
where $x_0 = ( k^2+b^2-k^2_{\rm F} )/(2k b)$. 
After performing the integration over $x$ and $k$, we find 
\begin{align}
{\cal M}_{+,xx}({\bm b}_0,k_{\rm F}) =  \frac{2}{3} \left (1-\frac{k_{\rm F}}{b} \right )^3 \Theta(k_{\rm F}-b)\Theta(2b-k_{\rm F})~.
\end{align}
For the other tensor elements we find 
\begin{eqnarray}\label{eq:Mc}
{\cal M}_{+,yy}({\bm b}_0,k_{\rm F}) = {\cal M}_{+,zz}({\bm b}_0,k_{\rm F}) &=&  -\frac{1}{2} {\cal M}_{+,xx}({\bm b}_0,k_{\rm F}),~~ {\cal M}_{+,\alpha\neq \beta}({\bm b}_0,k_{\rm F}) =0~. 
\end{eqnarray}
In summary, considering Eqs.~(\ref{eq:Mv}) and~(\ref{eq:Mc}), we have 
\begin{eqnarray}
{\cal M}_{{\rm cv},xx}({\bm b}_0,k_{\rm F}) = -2 {\cal M}_{{\rm cv},yy}({\bm b}_0,k_{\rm F}) &=& -2 {\cal M}_{{\rm cv},zz}({\bm b}_0,k_{\rm F}) = 
 \frac{2}{3} \left \{ \left (1-\frac{k_{\rm F}}{b} \right )^3 \Theta(k_{\rm F}-b)\Theta(2b-k_{\rm F}) - 1\right \}~.  
\end{eqnarray}
We now use a rotation around the $\hat{\bm k}_{y}$ axis in order to transform the vector ${\bm b}_0$ onto a generic vector ${\bm b} \equiv {\bm R}(\phi)  {\bm b}_0 = b [{\hat{\bm k}_{x}\cos(\phi) + \hat{\bm k}_{z}\sin(\phi)}] $ lying in the $\hat{\bm k}_{x}$-$\hat{\bm k}_{z}$ plane. Here, 
\begin{equation}
{\bm R}(\phi)=\begin{pmatrix} 
\cos(\phi)&0&-\sin(\phi)
\\
0&1&0
\\
\sin(\phi)&0&\cos(\phi)
\end{pmatrix}~.
\end{equation}
Accordingly, ${\bm {\mathcal M}}_{\rm cv}({\bm b},k_{\rm F}) = {\bm R}(\phi) {\bm {\mathcal M}}_{\rm cv}({\bm b}_0,k_{\rm F})  {\bm R}^{\rm T}(\phi)$. This immediately leads to Eq.~(11) in the main text.

\section{On the linear-response charge and node conductivities}\label{app:linear}
Starting from Eq.~(\ref{eq:linear-response-frequency-domain}), we can introduce anomalous {\it charge} and {\it node} linear-response conductivities: $ J^{(1), \rm c/n}_{\alpha}(\omega) = \sum_{\beta}\sigma^{(1), \rm c/n}_{\alpha\beta} {\cal E}_{\beta}(\omega) $. At zero temperature and zero doping, the anomalous first-order charge conductivity reads as following,
\begin{equation}
\sigma^{(1),{\rm c}}_{\alpha\beta} = -\frac{e^2}{\hbar}  \sum _{\gamma}\epsilon_{\alpha\beta\gamma} \int \frac{d^3{\bm k}}{(2\pi)^3} 
\sum^{N}_{i=1}   \Omega^{(\rchi_i)}_{-,\gamma}({\bm k})~,
\end{equation}
where $N$ indicates the total number of Weyl nodes. We can also introduce, in analogy to the valley conductivity in two-dimensional materials, the anomalous first-order ``node'' conductivity: 
\begin{equation}
\sigma^{(1),{\rm n}}_{\alpha\beta} = -\frac{e}{\hbar}  \sum _{\gamma}\epsilon_{\alpha\beta\gamma} \int \frac{d^3{\bm k}}{(2\pi)^3}  
\sum^{N}_{i=1} \rchi_i  \Omega^{(\rchi_i)}_{-,\gamma}({\bm k})~.
\end{equation}
For a single Weyl-Berry monopole located at ${\bm k} = {\bm b}$, 
\begin{equation}
{\bm \Omega}^{(\rchi)}_{n = \pm}({\bm k})=  n \rchi \frac{{\bm k}-{\bm b}}{2|{\bm k}-{\bm b}|^3}~,
\end{equation}
where $n=\pm$ indicates conduction/valence band states. We therefore need to evaluate the following integral:
\begin{equation}
\int d^3 {\bm k} \frac{{\bm k}-{\bm b}}{~|{\bm k}-{\bm b}|^3} =  \int^{\infty}_{0} k^2 dk \int^{1}_{-1} d\cos(\theta) \int^{2\pi}_{0} d\phi  \frac{(k\cos\theta-b)\hat{\bm b}+ k\sin\theta \hat{\bm \rho} }
{(k^2+b^2-2k b \cos\theta)^{\frac{3}{2}}}~,
\end{equation}
where $\hat{\bm \rho}\cdot \hat{\bm b}=0$ and $\int d\phi  \hat{\bm \rho} =0$ for symmetry. 
 We therefore have \cite{jackson} 
\begin{eqnarray}\label{eq:important_integral}
\int d^3 {\bm k} \frac{{\bm k}-{\bm b}}{~|{\bm k}-{\bm b}|^3} &=& 2\pi \hat{\bm b} \int^{\infty}_{0} k^2 dk \int^{1}_{-1} dx
 \frac{k x -b}{(k^2+b^2-2k b x)^{\frac{3}{2}}}
 \nonumber \\
 &=& - 2\pi \hat{\bm b} \frac{1}{b^2}  \int^{\infty}_{0} k^2 dk \left \{1+{\rm sign}[b-k] \right \} =- 4\pi \hat{\bm b} \frac{1}{b^2}  \int^{b}_{0} k^2 dk  = - \frac{4\pi}{3} {\bm b}~.
\end{eqnarray}
%
%We would like to note that the above result can be also concluded from Eq. 4.18 of  Ref. [\onlinecite{jackson}] that is also proved in the spherical coordinate. 
Note that it has been shown in Ref.~[\onlinecite{Goswami_prb_2013}] that the numerical prefactor changes by a factor of 3 if one performs the integral using a different regularization technique in cylindrical coordinates.
Using Eq.~(\ref{eq:important_integral}) and specializing to the case of a single Weyl-Berry dipole ${\bm D}=\sum_i \rchi_i {\bm b}_i$, like in Fig.~1(a) in the main text, we find 
 \begin{align}
 &\sigma^{(1),{\rm c}}_{\alpha\beta} = - \frac{e^2}{6\pi h} \sum_\gamma \epsilon_{\alpha\beta\gamma} D_{\gamma}~,
\\
&\sigma^{(1),{\rm n}}_{\alpha\beta} = 0~.
 \end{align}
For the case of the Weyl-Berry quadrupole in Fig.~1(b), we find $ \sigma^{(1),{\rm c}}_{\alpha\beta} =  \sigma^{(1),{\rm n}}_{\alpha\beta} =0$, because  $ \sum^4_{i=1} {\bm b}_i=\sum^4_{i=1} \rchi_i {\bm b}_i=0$~. 

In the linear-response regime, node currents vanish in both cases of systems with a Weyl-Berry dipole and quadrupole. In order to find a non-zero node current we need to go beyond the linear-response regime. 

\section{Anomalous circular photocurrent in strained TaAs-based WSMs}\label{app:TaAs}
The TaAs family of WSMs contains $N=24$ Weyl nodes distributed in three parallel crystal planes of the BZ. Each of these layers includes $N=8$ nodes, which can be decomposed in two independent quadrupoles with opposite sign, depicted by dashed red and black rectangles in Fig.~\ref{fig:TaAs}. 
The quadrupole tensor for the red (black) rectangle is denoted by ${\bm Q}_{i}$ ($-{\bm Q}'_{i}$)  where $i=-1,0,1$ refers to the bottom, middle, and top layer, respectively. 
In this regard, the total anomalous second-order charge conductivity is given by  
 \begin{align}\label{eq:sigma2c}
{\bm \sigma}^{(2),c}(\omega) = -  \frac{2e^3}{ h^2 \omega }  \sum_{i=-1,0,1}  \left \{  \frac{{\bm Q}_{i}}{d^2_{i}}-  \frac{{\bm Q}'_{i}}{{d'}^2_{i}} \right \}~, 
\end{align}
where $d_i=2\sqrt{b^2_{i,x}+b^2_{i,z}}$, $d'_i=2\sqrt{{b'}^2_{i,x}+{b'}^2_{i,z}}$, and 
\begin{equation}\label{eq:Q}
{\bm Q}_i =  4\rchi  b_{ i, x} b_{ i, z} \begin{pmatrix} 0&0&1\\0&0&0\\1&0&0\end{pmatrix}~,~{\bm Q}'_i =  4\rchi  b'_{ i, x} b'_{ i, z} \begin{pmatrix} 0&0&1\\0&0&0\\1&0&0\end{pmatrix}~.
\end{equation}
Here, $\rchi$ is the chirality of the node in the middle plane located at ${\bm b}= b_{0,x} \hat{\bm k}_x - b_{0,z} \hat{\bm k}_z$, which is denoted by a dashed yellow circle in Fig.~\ref{fig:TaAs}.
By inserting  Eq.~(\ref{eq:Q}) in Eq.~(\ref{eq:sigma2c}),  we obtain 
\begin{align}
{\bm \sigma}^{(2),c}(\omega) =-\rchi \frac{2 e^3}{h^2\omega} 
\left \{ 
 \frac{b_{0, x} b_{0, z}}{b^2_{0,x}+b^2_{0,z}} -  \frac{b'_{ 0, x} b'_{ 0, z}}{{b'}^2_{0,x}+{b'}^2_{0,z}}
-\sum_{i=\pm} \left [ \frac{b_{i, x} b_{i, z}}{b^2_{i,x}+b^2_{i,z}} -  \frac{b'_{i, x} b'_{i, z}}{{b'}^2_{i,x}+{b'}^2_{i,z}} \right ]
 \right \}
\begin{pmatrix} 0&0&1\\0&0&0\\1&0&0\end{pmatrix}~.
\end{align}
In a pristine sample of TaAs there is no strain and therefore the four-fold rotational symmetry of a tetragonal lattice implies $(b'_{i,x},b'_{i,z})=(b_{i,z},b_{i,x})$, yielding a vanishing anomalous second-order circular photocurrent. 
However, by straining the sample, one can break this rotational symmetry. Let us consider a uniaxial strain along the $\hat{\bm k}_{x}$ direction, ${\bm \varepsilon}={\rm diag}[\varepsilon,0,0]$. This modifies
the momentum-space vectors ${\bm b}_{i}$ and ${\bm b}'_{i}$: 
\begin{align}
b_{i,x} \to (1-\varepsilon) b_{i,x},~~b_{i,z} \to b_{i,z} ~,
\\
b'_{i,x} \to (1-\varepsilon) b_{i,z},~~b'_{i,z} \to b_{i,x}~. 
\end{align}
Consequently, 
\begin{align}
\frac{b_{i, x} b_{i, z}}{b^2_{i,x}+b^2_{i,z}} -  \frac{b'_{ i, x} b'_{ i, z}}{{b'}^2_{i,x}+{b'}^2_{i,z}} 
\to
\frac{(1-\varepsilon) b_{i,x} b_{i, z}}{(1-\varepsilon)^2 b^2_{i,x}+b^2_{i,z}} 
-  \frac{b_{ i, x} (1-\varepsilon) b_{i,z}}{ (1-\varepsilon)^2 b^2_{i,z} +b^2_{i,x}}  
\approx 2 \varepsilon \frac{b^2_{i,x}-b^2_{i,z}}{b^2_{i,x} + b^2_{i,z}}  \frac{b_{i,x}b_{i,z}}{b^2_{i,x} + b^2_{i,z}} +{\cal O}(\varepsilon^2)~.
\end{align}
Considering the definition of the quantities $g_i=  (b^2_{i,x}-b^2_{i,z}) / (b^2_{i,x} + b^2_{i,z})$, we finally reach the desired result   
\begin{align}
{\bm \sigma}^{(2), {\rm c}}(\omega)  \approx - \varepsilon \frac{4 e^3}{h^2\omega} 
\left \{ 
g_0\frac{{\bm Q}_0}{d^2_0}
-\sum_{i=\pm}  g_\pm\frac{{\bm Q}_\pm}{d^2_\pm}  
 \right \}~.
\end{align}
%
%\newpage
\end{widetext}

\begin{thebibliography}{99}
\bibitem{footnote}
	The first term on the right-hand side of Eq.~(\ref{eq:j_intra}) vanishes at first order in $\bm{\mathcal E}$ in general, 
	while we have checked that it vanishes at second order in $\bm{\mathcal E}$ for a WSM.
\bibitem{footnote_dyadic}
For the case of $\hat{\bm n}= \hat{\bm k}_y$ and  $\hat{Q}_{ij}= \delta_{i x}\delta_{j z} +\delta_{i z}\delta_{j x}$, we have 
$\hat{\bm n} \times \hat{\bm Q} = \sum_{ij} \hat{Q}_{ij} (\hat{\bm k}_y\times\hat{\bm k}_i) \hat{\bm k}_j =  \hat{\bm k}_x\hat{\bm k}_x-\hat{\bm k}_z\hat{\bm k}_z$ which is a diagonal tensor.
\end{thebibliography}
 \end{document}